# AN ESTIMATING EQUATIONS APPROACH TO FITTING LATENT EXPOSURE MODELS WITH LONGITUDINAL HEALTH OUTCOMES[1]

By Brisa N. Sánchez, Esben Budtz-Jørgensen and Louise M. Ryan

*University of Michigan, University of Copenhagen and Harvard University*

The analysis of data arising from environmental health studies which collect a large number of measures of exposure can benefit from using latent variable models to summarize exposure information. However, difficulties with estimation of model parameters may arise since existing fitting procedures for linear latent variable models require correctly specified residual variance structures for unbiased estimation of regression parameters quantifying the association between (latent) exposure and health outcomes. We propose an estimating equations approach for latent exposure models with longitudinal health outcomes which is robust to misspecification of the outcome variance. We show that compared to maximum likelihood, the loss of efficiency of the proposed method is relatively small when the model is correctly specified. The proposed equations formalize the ad-hoc regression on factor scores procedure, and generalize regression calibration. We propose two weighting schemes for the equations, and compare their efficiency. We apply this method to a study of the effects of in-utero lead exposure on child development.

**1. Introduction.** The association between child lead exposure and neurodevelopment has been widely studied. Initially research focused on describing the cross-sectional relationship between lead exposure measured by child's blood lead concentration and mental development [e.g., Fulton et al. (1987), Hatzakis et al. (1989)], and later on the association between development and prenatal exposure measured by the concentration of lead in

Received December 2007; revised November 2008.
[1]Supported by a Predoctoral Fellowship from the Howard Hughes Medical Institute (HHMI) to the first author. Supported also in part by a grant from the American Chemistry Council (ACC 2843).

umbilical cord blood [e.g., Bellinger (1989)]. More recently, lead concentrations in the maternal skeleton and in plasma (a component of whole blood) emerged as novel biomarkers of prenatal exposure, and studies have shown inverse associations with mental developmental at 24 months [e.g., Gomaa et al. (2002); Hu et al. (2006)]. Currently, one focus of this area of research is on the time windows during pregnancy when the developing fetus is more vulnerable to exposure [e.g., Schnaas et al. (2006)].

Studying the issue of time windows of vulnerability to fetal lead exposure is difficult since direct measures of fetal exposure are not feasible. Cord blood lead levels have been used as a measure of fetal exposure, but only reflect the third trimester of exposure because the half-life of lead in blood is approximately thirty to 45 days [Hu et al. (1998)]. Exposures earlier during pregnancy, for example, the first trimester, may be more important as the developing brain may be more susceptible to neurotoxicants during this period [Mendola et al. (2002)].

1.1. *ELEMENT study in Mexico City.* The Early Life Exposures in Mexico City to Neuro-Toxicants (ELEMENT) study recruited prospective mothers, at or before conception, to address the question of windows of vulnerability to lead exposure. Women were followed during pregnancy to assess their exposure to lead. Their children were followed after birth to assess their development and lead exposure using, respectively, the mental development index (MDI) of the Bayley's scale of mental development [Bayley (1993)] and blood lead levels at 3, 12, 18, 24, 30 and 36 months of age. Various measures of exposure (lead concentrations in whole blood and plasma) were collected on the mother during each trimester of pregnancy and at 1 month postpartum; for a small group of mothers, all this information was also captured before conception. Measures of lead concentration in blood and plasma during pregnancy are the closest surrogate measures of fetal exposure. Lead stored in maternal bone leaches out into blood and is thus considered an important source of exposure to the fetus, and an important predictor of plasma and blood lead levels. Another predictor of blood and plasma levels is the rate at which bone resorbs (natural bone remodeling process). Urinary cross-linked N-telopeptive (NTx), a measure of bone resorption in units of bone collagen equivalents, was also collected. Other information, such as maternal age and IQ, and the use of lead-contaminated ceramics (days/week) was also collected. Table 1 lists sample characteristics for the outcome as well as covariates, and thirteen exposure surrogates; the sample consists of 341 mother–child pairs. To be included in the analysis we present, the mother had to have measurements on at least one of the surrogate measurements of fetal exposure, and have completed an IQ test. The children in the sample completed at least one of six assessments of Bayley's MDI, and had a concurrent blood lead measurement at the time of each MDI assessment.

Table 1
*ELEMENT study in Mexico City: sample characteristics*

|  | Time | $N$ | Mean | StDev |
|---|---|---|---|---|
| **Prenatal exposure surrogates** | | | | |
| Mother's $\log_2$ Plasma lead | BP | 11 | 4.1 | 1.05 |
|  | T1 | 153 | 3.8 | 0.96 |
|  | T2 | 169 | 3.4 | 0.86 |
|  | T3 | 157 | 3.5 | 0.79 |
| **Mother's whole blood lead** | | | | |
| Laboratory 1 | BP | 11 | 8.8 | 9.0 |
|  | T1 | 155 | 6.9 | 4.7 |
|  | T2 | 173 | 6.2 | 3.1 |
|  | T3 | 159 | 6.7 | 3.4 |
| Laboratory 2 | BP | 29 | 8.0 | 5.8 |
|  | T1 | 172 | 7.6 | 4.6 |
|  | T2 | 198 | 6.6 | 3.2 |
|  | T3 | 304 | 6.9 | 3.6 |
| Child's $\log_2$ Cord blood lead | Birth | 238 | 2.1 | 0.9 |
| **Post natal exposure** | | | | |
| Child's blood lead | 3mpp | 300 | 3.8 | 1.9 |
|  | 12mpp | 298 | 4.9 | 2.9 |
|  | 18mpp | 321 | 6.8 | 3.6 |
|  | 24mpp | 318 | 4.8 | 3.4 |
|  | 30mpp | 248 | 6.5 | 3.6 |
|  | 36mpp | 280 | 6.9 | 3.7 |
| **Health outcome** | | | | |
| Child's MDI | 3mpp | 323 | 94.2 | 5.7 |
|  | 12mpp | 323 | 95.4 | 9.1 |
|  | 18mpp | 321 | 91.1 | 8.6 |
|  | 24mpp | 318 | 91.4 | 11.3 |
|  | 30mpp | 248 | 92.9 | 8.5 |
|  | 36mpp | 280 | 94.1 | 8.5 |
| **Mother's bone measures** | | | | |
| Tibia lead | BP | 19 | 9.4 | 12.1 |
| Patella lead | BP | 23 | 13.2 | 10.9 |
| Tibia lead | 1mpp | 279 | 7.8 | 9.6 |
| Patella lead | 1mpp | 335 | 10.7 | 10.7 |
| NTx | T1 | 95 | 6.1 | 0.74 |
|  | T2 | 127 | 6.5 | 0.75 |
|  | T3 | 137 | 7.0 | 0.59 |

Details on the laboratory procedures to obtain these measures are reported elsewhere [e.g., LaMadrid-Figueroa et al. (2006); Tellez-Rojo et al. (2004)]. Briefly, bone concentrations ($\mu$g Pb/g bone mineral) are obtained

TABLE 1
*(Continued)*

|  | Time | $N$ | Mean | StDev |
|---|---|---|---|---|
| Covariates |  |  |  |  |
| Mother's IQ | 24mpp | 341 | 89.6 | 18.6 |
| Age | Scr | 341 | 25.9 | 5.1 |
| % of life in Mexico City | Scr | 341 | 89.9 | 27.3 |
| Weekly ceramics use | Scr | 341 | 0.60 | 1.55 |
| Child's gender | Birth | 341 | 0.5 |  |

Abbreviations: BP = Before Pregnancy; T$j$ = Trimester $j$; $x$mpp = $x$ months post partum; Scr = Screening.

with a K-X ray instrument (similar to a regular X-ray, but emits lower radiation). A urine sample and two blood samples were taken at each prenatal visit. One blood sample was sent to "Laboratory 1," where whole blood lead concentration was measured ($\mu$g Pb/dL). The other blood sample and the urine were sent to "Laboratory 2." The samples were processed to obtain the concentration of lead in plasma ($\mu$g Pb/dL) as well as a second measure of lead in whole blood, and the urinary NTx measure, respectively.

TABLE 2
*Estimated associations$^\sharp$ between mental development and prenatal and postnatal lead exposure, using available surrogates for prenatal exposure*

|  |  | Prenatal exposure | | | Postnatal exposure | | |
|---|---|---|---|---|---|---|---|
| Surrogate | N | $\widehat{\beta}_1$ | s.e. | $p$-value | $\widehat{\beta}_2$ | s.e. | $p$-value |
| $\log_2$ Plasma lead $_{t=1}$ | 153 | −0.423 | 0.482 | 0.380 | −0.111 | 0.144 | 0.442 |
| $\log_2$ Plasma lead $_{t=2}$ | 169 | −0.623 | 0.509 | 0.221 | −0.073 | 0.145 | 0.616 |
| $\log_2$ Plasma lead $_{t=3}$ | 157 | −0.429 | 0.516 | 0.406 | −0.122 | 0.145 | 0.403 |
| Mother's whole blood lead |  |  |  |  |  |  |  |
|  Laboratory 1$_{t=1}$ | 155 | −0.834 | 0.366 | 0.023 | −0.093 | 0.140 | 0.508 |
|  Laboratory 1$_{t=2}$ | 173 | −1.201 | 0.417 | 0.004 | −0.023 | 0.141 | 0.872 |
|  Laboratory 1$_{t=3}$ | 159 | −0.960 | 0.494 | 0.052 | −0.063 | 0.145 | 0.664 |
|  Laboratory 2$_{t=1}$ | 172 | −0.935 | 0.405 | 0.021 | −0.111 | 0.139 | 0.426 |
|  Laboratory 2$_{t=2}$ | 198 | −1.048 | 0.377 | 0.005 | −0.041 | 0.136 | 0.762 |
|  Laboratory 2$_{t=3}$ | 304 | −0.485 | 0.324 | 0.135 | −0.162 | 0.113 | 0.154 |
| $\log_2$ Cord blood lead | 238 | −0.223 | 0.422 | 0.598 | −0.312 | 0.124 | 0.012 |

$^\sharp$From regression models for longitudinal data estimated with generalized estimating equations, assuming exchangeable correlation structure, and adjusted for maternal age and IQ, child's gender, child's age using indicator variables for each time point, child's blood lead concentration and gender by time interactions.

One of primary goals of the study was to obtain an unbiased measure of association between development and fetal exposure during the first trimester of pregnancy, while adjusting for postnatal (childhood) exposures. Typically, regression analysis is used to describe such association, while adjusting for other factors such as maternal age and IQ, and child's blood lead concentration at the time of the MDI measurement. Because fetal exposure is not directly observed, any of up to thirteen lead concentrations could serve as a marker of this exposure. Table 2 gives the regression coefficient for each of the prenatal exposure surrogates. For ease of comparison across multiple surrogates, the coefficients are expressed in units of $\log_2$-plasma lead during the first trimester (i.e., we first standardize each surrogate to have mean zero and unit variance, multiply by the standard deviation of first trimester $\log_2$-plasma lead, and estimate the model using the transformed surrogate).

Interpreting the results in Table 2 gives rise to various statistical concerns. One concern is multiple testing, since an immediate attempt at interpretation would be to compare the significance of each coefficient against a predetermined significance level. Using a Bonferroni correction to correct for multiple testing would lead to no "significant" findings. Another concern is that the models use only available cases, such that not all regressions include the same mother–child pairs. However, mothers with available data for these regressions are older, more likely to use contaminated ceramics, and have higher IQ. Further, the regression coefficients may be biased since this approach does not account for measurement error in the surrogate measures of exposure. Last, including more than one (or at most a few) prenatal exposure biomarkers in the same model leads to problems with collinearity, as the correlations among prenatal exposure biomarkers are high, ranging from 0.52 to 0.9. Thus, although this data collection provides a wealth of exposure information, it also gives rise to methodological issues when analyzing the association between exposure and health outcomes.

1.2. *Latent variables: a framework for modeling complex exposure data and implications for ELEMENT study.* Latent variable modeling, which has been extensively used in social science research [Bollen (1989)], is an increasingly popular analysis tool in medical research and environmental epidemiology studies involving complex exposure data [Budtz-Jørgensen et al. (2002); Nikolov et al. (2006); Rabe-Hesketh and Skrondal (2008)]. Latent variable models provide a framework to reduce the dimensionality of the exposure data and incorporate information regarding the underlying biological relationships between the exposure measurements. In this modeling framework, multiple measures of an exposure can be viewed as surrogates for a true but unobserved exposure, reducing the dimensionality of predictor variables. In the in-utero lead study, the various measured lead concentrations can be viewed as characterizing latent fetal exposure that changes with time [e.g., Roy and Lin (2000)]. The child's blood lead concentrations can be viewed as indirect measures of an overall postnatal exposure. In the next section we propose a latent variable model which succinctly describes the association between mental development and the in-utero and post-natal lead exposure data, and is more parsimonious compared to multiple regression analysis.

Although the advantages of latent variable models are clear, some of the disadvantages include their lack of robustness to model misspecification and some disagreements about parameter estimation. Latent variable models are susceptible to the assumptions imposed on the covariance structure of residual errors. That is, incorrect variance specification leads to biased estimation of the parameters of primary interest, namely those describing the association between the exposure and outcome [Reddy (1992); Hoogland and Boomsma (1998); Sammel and Ryan (2002)]. In the lead study, the statistical significance and magnitude of the latent exposure coefficient obtained via maximum likelihood estimation depends heavily on the assumed covariance structure for the MDI residuals (Table 3). This provides empirical evidence of the strong dependence of the regression coefficients on the covariance structure for the outcome residuals and related bias. In a simulation study we quantify the magnitude of the bias and show that it is not in a consistent direction. While methods that relax classic distributional assumptions (e.g., normality) on the error terms have been developed [Browne (1984); Arminger and Schoenberg (1989)], the more challenging problem of relaxing correct specification of residuals' covariance matrices remains an open problem.

A related area of discussion regarding parameter estimation involves models where surrogates of exposure can be modeled separately from the model relating the latent exposure to the health outcome. Joint estimation of the model for the exposure measurements and the regression model between the latent exposure is advocated by many. The primary arguments in favor of joint estimation are that it increases efficiency and eliminates possible biases from two step approaches to estimation [Bartholomew (1981); Bollen (1989); Iwata (1992); Wall and Li (2003)]. However, joint estimation can make the models more susceptible to model misspecification [Hoogland and Boomsma (1998)]. In the fetal lead exposure example, we were particularly concerned about obtaining biased effects of lead exposure due to misspecified covariance assumptions on the longitudinal health outcome (Bayley's MDI); the estimated effect varies greatly depending on the covariance assumption. Hence, we sought an alternative estimation approach which is robust to misspecification, but does not lose too much efficiency.

We propose an estimating equations approach for models with latent exposures and longitudinal outcomes that is robust to misspecification of the conditional variance of the longitudinal outcomes given the true exposure.

TABLE 3
*Estimated effects*[†] *of latent trimester 1 exposure and postnatal exposure on child mental development based on the model from Figure [1]; results from various fitting procedures*

| Method | Covariance structure | Prenatal exposure | | | Postnatal exposure | | |
|---|---|---|---|---|---|---|---|
| | | $\widehat{\beta}_1$ | s.e. | $p$-value | $\widehat{\beta}_2$ | s.e. | $p$-value |
| MLE | Independence | −2.4600 | 0.5280 | 0.000 | −0.7730 | 0.5520 | 0.081 |
| | CS | −1.0370 | 0.5730 | 0.035 | −0.8480 | 0.6430 | 0.094 |
| | CSH | −0.9420 | 0.5390 | 0.040 | −0.7500 | 0.6010 | 0.106 |
| | Unstructured | −0.6350 | 0.4340 | 0.072 | −0.2110 | 0.4870 | 0.333 |
| GEE | Independence | | | | | | |
| | $(\beta_1^*, \beta_2^*) = (0,0)$ | −1.0231 | 0.5745 | 0.037 | −0.9719 | 0.5484 | 0.038 |
| | $(\beta_1^*, \beta_2^*) = (-1.0, -0.8)$ | −1.0290 | 0.5729 | 0.036 | −0.9623 | 0.5462 | 0.039 |
| | $(\beta_1^*, \beta_2^*)$ not fixed | −1.0400 | 0.5733 | 0.035 | −0.9650 | 0.5464 | 0.039 |
| GEE | Exchangeable | | | | | | |
| | $(\beta_1^*, \beta_2^*) = (0,0)$ | −0.9924 | 0.5604 | 0.038 | −0.8767 | 0.5535 | 0.057 |
| | $(\beta_1^*, \beta_2^*) = (-1.0, -0.8)$ | −0.9941 | 0.5598 | 0.038 | −0.8733 | 0.5526 | 0.057 |
| | $(\beta_1^*, \beta_2^*)$ not fixed | −0.9967 | 0.5599 | 0.038 | −0.8757 | 0.5527 | 0.057 |
| GEE | Exchangeable♯ | | | | | | |
| | $\beta_1^* = 0$ | −1.1021 | 0.4911 | 0.012 | −0.1389 | 0.1074 | 0.098 |
| | $\beta_1^* = -1.0$ | −1.1016 | 0.4908 | 0.012 | −0.1392 | 0.1073 | 0.097 |

[†]Adjusted for maternal age and IQ, child's gender, child's age using indicator variables for each time point and gender by time interactions.

♯ Model uses observed child's blood lead concentration as measure of postnatal exposure.

The proposed approach can be viewed as a generalization of regression calibration [Carroll et al. (2006)], and formalizes an ad-hoc procedure popular among psychometricians, namely, regression on factor scores [Tucker (1971); Skrondal and Laake (2001)]. It loses little efficiency compared to maximum likelihood estimation when model assumptions are met. In addition, we show that, under certain study designs, our proposed estimating equations approach can yield estimates that are more efficient than those from regression calibration. The proposed approach can easily accommodate studies with many patterns of missing data among the exposure measurements.

The organization of this paper is as follows. In Section 2 we present more details about the study of in-utero lead exposure and neurodevelopment in children, propose a latent variable model for the data, and use the example to introduce many of the latent variable concepts. In Section 3 we develop a more general model for longitudinal responses where predictors of interest are latent. This more general specification allows us to more succinctly describe estimation methods, and bias and efficiency issues in Sections 4 and 5, respectively. Specifically, in Section 4 we review maximum likelihood es-timation, introduce the robust estimating equations approach, and discuss how the proposed approach generalizes factor-score regression and regression calibration. Section 5 compares maximum likelihood estimation, regression calibration and the proposed approach in terms of bias, mean squared error and variance of the estimated exposure effect. Finally, Section 6 discusses the conclusions and implications of the estimation approach proposed herein.

**2. Latent variable model for latent lead exposure, and neurodevelopment.** Given the large number of exposure measurements available, we develop a way to synthesize the exposure information using a latent variable model. We think of the overall model as two separate pieces: a model for the exposures, "the exposure model," and a model that links the latent exposures to the mental development outcome, "the outcome model." In synthesizing the exposure information, we take into account the longitudinal aspect of the exposures, as well as the biological relationships between the various exposure biomarkers. The model is detailed below algebraically, and is also shown graphically in Figure 1.

The exposure part of the model takes into account the longitudinal assessment of the exposure biomarkers by positing the existence of a latent exposure at each trimester of pregnancy. The latent exposure at each trimester is assumed to be indirectly measured by plasma and blood lead concentrations. That is, letting $U_{it}$ represent the latent exposure for individual $i$ at each trimester of pregnancy $t$, $t = 0, 1, 2, 3$ ($t = 0$ refers to before pregnancy), we propose

$$
\begin{aligned}
&X_{i1t} = U_{it} + \delta_{i1t}, && \text{model for plasma lead,} \\
&X_{i2t} = \nu_{2t} + \lambda_2 U_{it} + \delta_{i2t}, && \text{model for blood lead (Laboratory 1),} \\
&X_{i3t} = \nu_{3t} + \lambda_3 U_{it} + \delta_{i3t}, && \text{model for blood lead (Laboratory 2),} \\
&X_{i43} = \nu_{43} + \lambda_4 U_{i3} + \delta_{i43}, && \text{model for cord blood,}
\end{aligned}
\tag{1}
$$

where $\delta$'s are zero-mean measurement errors. Given that $U_{it}$'s are not observed, they do not have a natural location and scale. In this model, however, the latent lead exposure is assumed to have the same mean and the same units as lead concentration in plasma. This information in conveyed by the equation $X_{i1t} = U_{it} + \delta_{i1t}$. The $\nu$'s and $\lambda$'s in the other equations shift the location and translate units between the various measurements and the latent exposures.

Latent variable models typically assume that surrogate measures $X$ are conditionally independent of each other given the latent variables, for example, $U_t$. In other words, the covariance matrix for the $\delta$'s is assumed to be diagonal. However, after modeling the mean of the blood lead levels, correlations among the residuals within laboratory may exist. We model the correlation of the errors $\delta_{2t}, t = 0, \ldots, 3$, across time as using an autocorrelation structure of order 1: $\text{cor}(\delta_{20}, \delta_{21}) = \text{cor}(\delta_{21}, \delta_{22}) = \text{cor}(\delta_{22}, \delta_{23}) = \rho_1$, and similarly for $\delta_{3t}, t = 0, \ldots, 3$. However, $\delta_{2t}$ is assumed independent from $\delta_{3t'}$ for all time points $t, t'$ (different laboratories). In Figure 1 these correlations are depicted with double headed arrows between the respective measures.

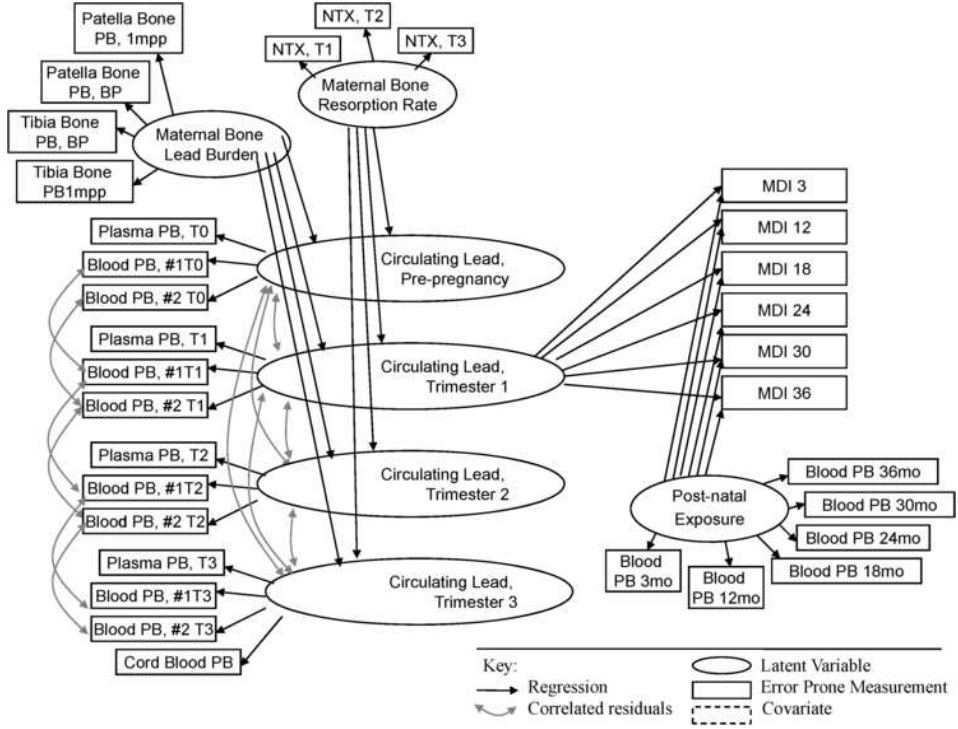

FIG. 1. *Path diagram showing relationships between exposure surrogates, latent exposures and mental development index. Covariates are not shown in the figure. However, all latent variables and Bayley's MDI are regressed on maternal age. Additionally, bone lead burden is regressed on maternal percent of life in Mexico City; circulating lead levels are regressed on weekly leaded ceramic use; Bayley's MDI additionally regressed on maternal IQ and child's gender, indicator variables for measurement occasion, and gender by occasion interaction.*

Bone lead concentration was measured from the patella and tibia bones before and after pregnancy, for a total of up to four measures of bone lead from each mother (two before conception and two more at one month post partum), and up to three measures of bone resorption rates were collected. Bone resorption rates, measured at trimester $t = 1, 2, 3$, are modeled using $X_{t4} = U_4 + k_1 t + \varepsilon_{t4}$, where $U_4$ represents the mother's intrinsic resorption rates, and $k_1$ is a fixed time effect. The model for bone lead concentrations

is

$$X_{i15} = U_{i5} + \delta_{i15}, \qquad \text{patella lead, before pregnancy,}$$
$$X_{i25} = \nu_{25} + \lambda_{25}U_{i5} + \delta_{i25}, \qquad \text{tibia lead, before pregnancy,}$$
$$X_{i35} = \nu_{35} + U_{i5} + \delta_{i35}, \qquad \text{patella lead, 1 mo. post partum,}$$
$$X_{i45} = \nu_{45} + \lambda_{25}U_{i5} + \delta_{i45}, \qquad \text{tibia lead, 1 mo. post partum,}$$

which assumes that bone lead burden, $U_5$, is measured in units of patella lead concentration. The population average for bone lead burden is assumed to be equal to the average observed for patella lead before pregnancy, and changes by $\nu_{35}$ after pregnancy.

The next step in defining the exposure model is to relate latent variables to each other and to covariates which may predict the latent exposure. The purpose of this model is to be able to borrow exposure information across individuals who may have similar covariates or exposure levels. This is helpful in characterizing first trimester fetal exposure when mothers were not interviewed at the earlier trimesters of pregnancy.

Circulating lead levels change with time and are affected by maternal lead burden and the rate at which bone resorbs. Thus, we model them as $U_t = \alpha_0 + \gamma_{1,1}U_4 + \gamma_{1,2}U_5 + \gamma_{2,1}s(t) + \gamma_{2,2}W_1 + \xi_t$, where $s(t) = t$, for $t = 0, 1, 2$, and $s(3) = 2$; $W_1$ is use of contaminated ceramics, and $\text{var}[(\xi_0, \ldots, \xi_3)^T]$ is unstructured to avoid possible residual error covariance misspecification in the exposure model. The effects of bone resorption rates, $U_4$, and bone lead burden, $U_5$, on circulating lead are given by $\gamma_{1,1}$ and $\gamma_{1,2}$, respectively. Finally, we model the effects of covariates on bone lead burden and bone resorption rates: $U_4 = \alpha_4 + \gamma_{2,3}W_2 + \xi_4$ and $U_5 = \alpha_5 + \gamma_{2,4}W_2 + \gamma_{2,5}W_3 + \xi_5$, where $W_2, W_3$ are maternal age and percent of life lived in Mexico City. Increases in maternal age are associated with increased bone lead concentration and bone resorption, while larger percent of life lived in Mexico City (i.e., longer exposure time) is associated with increased bone lead burden. With this we have arrived at a model for prenatal exposures.

The postnatal environment can also be described using a latent variable assumed to be indirectly observed through the child's blood lead measurements at the time of the MDI assessments. Denoting $j$ as the measurement occasion for the child's MDI and blood lead, the model for $X_{ij6}$, the blood lead concentration at the $j$th occasion, is $X_{ij6} = \nu_{j6} + \lambda_{j6}U_{i6} + \delta_{ij6}$, where $U_{i6}$ represents postnatal exposure. We fix the location and scale of the postnatal exposure variable to that of $X_{i16}$. In Figure 1 the postnatal exposure variable is assumed to be independent from prenatal circulating lead exposure (no double headed arrows connecting them). This assumption slightly simplifies the figure, and has little effect on the estimated parameters of interest (e.g., <10% change in the parameters in Table 3).

Finally, we arrive at the "outcome model" linking the exposure model to the longitudinal health outcome. Denoting $Y_{ij}$ as the $j$th outcome for the $i$th child, the longitudinal outcomes model is $Y_{ij} = \beta_{0j} + \beta_1 U_{i1} + \beta_2 U_{i6} + \kappa_1^T Z_{ij} + \varepsilon_{ij}$, where $Z_{ij}$ is mother's age and IQ, child's gender, and occasion by gender interactions. The coefficient $\beta_1$ represents the association between latent first trimester exposure and child development. The contribution of the postnatal environment to MDI scores is modeled by $\beta_2 U_{i6}$; alternatively, $U_{i6}$ could be replaced by the observed blood lead concentration at the $j$th occasion $X_{ij6}$. For direct comparability of the estimated $\beta_2$, in this alternative model we use $\beta_2 X_{ij6}^s$, where $X_{ij6}^s$ is a version of $X_{ij6}$ scaled to have variance equal to $\text{var}(X_{i16})$. Residuals $\varepsilon_i = (\varepsilon_{i1}, \ldots, \varepsilon_{in_i})$ are assumed to be independent of $U_{i1}$, $U_{i6}$ and $Z_{ij}$, but can have various types of correlation structures among themselves. In Figure 1 the variance matrix for $\varepsilon_i$ is assumed to be diagonal, since there are no correlation arrows (double-headed) connecting the outcomes $Y_{ij}$ to each other. We explore the consequences of misspecifying this covariance assumption.

In the next sections we will discuss the details of the estimating approaches for this type of model, but at this time we give estimates of the regression coefficients obtained under various approaches. Table 3 shows the parameter estimates using maximum likelihood estimation (MLE) for model parameters under various choices of the correlation matrix for $\varepsilon_i$. The MLE estimate of $\beta_1$ more than doubles when the assumed correlation structure for the outcome is independence in comparison to compound symmetry. In some instances the maximum likelihood estimator is highly significant ($p < 0.001$), and in others not significant ($p = 0.07$). Variability in the estimated effect of postnatal exposure is also observed, although not to the same degree. The variability in the estimated exposure coefficients and in the inferences obtained under different covariance structures raises concern about the robustness of the MLE approach. Table 3 also shows estimates obtained via estimating equations approaches and regression calibration described in later sections. For prenatal exposure, the estimates and $p$-values from these alternative approaches are stable at about 1.0 point decline in MDI for a doubling of plasma lead concentration, and $p = 0.037$, respectively.

Estimates of other parameters, for example, maternal IQ and age coefficients, are stable across the MLE and estimating equation approaches. This supports the fact that MLE estimation gives biased estimates only for the latent variable coefficients when the outcome's covariance structure is misspecified (Section 4.1). The estimated increase in MDI is 0.54 points (95% CI: 0.01, 1.1) for a 5-year increase in maternal age and 0.67 points (95% CI: 0.35, 1.02) for a 10 point increase in maternal IQ. The estimated association between MDI and gender is also stable across estimation approaches, although it significantly varies across measurement occasion ($p < 0.001$). The largest gender difference occurs at 24 months, where girls score an average of 4.8 points higher than boys (95% CI: 2.2, 7.5).

Table 3 also shows the results from modeling the postnatal environment by replacing $U_{i6}$ by the observed child's blood lead concentration at the $j$th occasion. As can be seen, the parameter estimate is greatly attenuated toward zero, as would be expected under the assumption that child's blood lead is a surrogate measure (with error) of the postnatal environment [Carroll et al. (2006)]. However, the association with maternal age also changes by nearly 10% to 0.60 (95% CI: 0.04, 1.16), as might be expected given that the effect of the measurement error in one variable may extend to other covariates [Budtz-Jørgensen et al. (2003)].

**3. More general framework: latent exposures model with longitudinal responses.** Before we discuss the estimation approaches used to arrive at the estimates in Table 3, we write a more general framework for the model discussed in the previous section. Let $Y_i = (Y_{i1}, Y_{i2}, \ldots, Y_{in_i})^T$ represent a vector of $n_i$ continuous responses taken on the $i$th of $N$ subjects at occasions $j = 1, 2, \ldots, n_i$. These responses may depend on $l$ latent exposures $U_i = (U_{i1}, \ldots, U_{il})^T$, which are indirectly measured by $p$ observed surrogates $X_i = (X_{i1}, \ldots, X_{ip})^T$. In the example $l = 6$, but we were interested only in the effect of $U_{i1}$ and $U_{i6}$ on $Y_i$. Covariate data are denoted by $W_i = (W_{i1}, \ldots, W_{ir})^T$ and $Z_i = (Z_{i1}, \ldots, Z_{in_i})^T$. Covariates $Z_{ij}$ are $q \times 1$ vectors assumed to be measured concurrently with the $j$th repeated outcome, $Y_{ij}$, and may include variables to model time effects. Covariates $W_i$ are ascertained concurrently with $X_i$.

The exposure model can then be succinctly written as

$$X_i = \nu + \Lambda U_i + K W_i + \delta_i, \tag{2}$$

with the latent predictors $U_i$ depending on fixed covariates and other latent variables

$$U_i = \alpha + \Gamma_1 U_i + \Gamma_2 W_i + \xi_i, \tag{3}$$

and the outcome model as

$$Y_{ij} = \beta_0 + \beta^T U_i + \kappa^T Z_{ij} + \varepsilon_{ij}. \tag{4}$$

In (2) the $(p \times 1)$ vector of errors $\delta_i$ represents measurement variability, including measurement error; we assume $E(\delta_i | U_i, Z_i, W_i) = 0$ and $\text{var}(\delta_i | U_i, Z_i, W_i) = \Omega_\delta$. In (3) $\Gamma_1$ is an $l \times l$ matrix with all diagonal elements equal to zero, and we assume that $(I_l - \Gamma_1)$ is invertible, where $I_l$ is an identity matrix of the same dimension. In the lead example, the only nonzero entries of $\Gamma_1$ are those corresponding to $\gamma_{1,1}$ and $\gamma_{1,2}$, the effects of bone lead concentration and bone resorption rate on circulating lead levels. We suppose $E(\xi_i | Z_i, W_i) = 0$ and $\text{var}(\xi_i | Z_i, W_i) = \Psi$, and that $\xi_i$ is independent from $\delta_i$. The dimensions of the parameter matrices $\nu$, $\Lambda_2$ and $K$ are

$p \times 1$, $p \times l$ and $p \times r$, respectively; and those of $\alpha$ and $\Gamma_2$ are $l \times 1$ and $l \times r$, respectively.

In the outcome model, $E(\varepsilon_i | U_i, Z_i, W_i) = 0$, and $\text{var}(\varepsilon_i | U_i, Z_i, W_i) = \Omega_{\varepsilon_i}$, where the $n_i \times n_i$ matrix $\Omega_{\varepsilon_i}$ can vary with $i$ [e.g., for random effects with variance $\Delta$, $\Omega_{\varepsilon_i} = Z_i \Delta Z_i^T + \sigma^2 I_{n_i}$, where $I_{n_i}$ represents an $n_i \times n_i$ identity matrix, Laird and Ware (1982)]; in the sequel we may drop the subscript $i$ in this matrix for ease of exposition. For notational convenience, we assume that if any covariates $W_i$ are to be included in the outcome model, then they are also included in $Z_i$. Further, it is assumed that $\varepsilon_i$, $\delta_i$ and $\xi_i$ are mutually independent; typically these errors are also assumed to be normally distributed. Finally, as it will be further discussed in Section 4, we allow subjects to have missing data on some of the surrogate measurements, $X_i$. We hereafter refer to $(\beta_0, \beta^T, \kappa^T)$ as conditional mean parameters, and note that $\beta$ is the parameter of primary interest.

Extensive discussions on the interpretation and identifiability of other parameters in (2)–(4) appear in the literature [e.g., Bollen (1989); Skrondal and Rabe-Hesketh (2004); Sánchez et al. (2005)], thus, we only make a few remarks. First, note that the equation describing the relationship between surrogates and latent variables (2) also includes a covariate term. The matrix $K$ is typically sparse, with a few nonzero elements that allow for item bias [Beck (1982)]. Item bias consists of differential effects of covariates on particular surrogates that go beyond the effect of the covariate on the (parent) latent variable. In psychometrics, where these models have enjoyed much use, a typical example of item bias is differential responses by gender on a specific item (surrogate). In the context of the lead example, the surrogates are biomarkers of exposure, such that demographic characteristics may not necessarily affect a particular surrogate differently from other surrogates. However, one possible item bias covariate might be a genetic variant. For instance, it is hypothesized that the ALAD genotype changes the affinity of lead to red blood cells [Bergdahl et al. (1998)]. Thus, people with the ALAD variant would have different average lead levels in red blood cells compared to ALAD-wildtypes, but the concentrations of plasma lead may not necessarily differ.

Second, the matrix $\Lambda$ will typically have a pre-defined pattern of zeroes and ones to reflect knowledge about the relationships between surrogates and latent variables and defining the scales of the latent variables. Given the patterns of zeroes, (2) can be thought of as confirmatory factor analysis, which avoids the identifiability concerns typically associated with factor analysis [Bollen (1989); Jolliffe (1998)].

**4. Estimation.** We now discuss maximum likelihood estimation and the proposed estimating equations approach using the more general framework for the model discussed in Section 3. We write score equations for the model to study how invalid covariance assumptions induce bias in the estimate for $\beta$. We then introduce estimating equations to eliminate the bias, generalize regression calibration, and formalize regression on factor scores.

We begin by introducing notation common to all approaches. Let $\theta$ denote all model parameters. Because it will be helpful in efficiency calculations later, we partition $\theta$ into $\theta^T = (\theta_1^T, \theta_2^T, \theta_3^T)$, where $\theta_1^T = (\beta_0, \beta^T, \kappa^T)$ are parameters for the conditional mean in the outcome model, $\theta_2$ parameterizes the variance matrix of the outcome given the latent exposure, that is, $\Omega_\varepsilon = \Omega_\varepsilon(\theta_2)$, and $\theta_3$ is a vector containing the exposure model parameters, that is, those that parameterize $\nu, \Lambda, K, \Omega_\delta, \alpha, \Gamma_1, \Gamma_2$ and $\Psi$.

The estimation approaches utilize several marginal and conditional moments, which we now define. Given covariates $W_i$ for individual $i$, the mean and variance of the latent exposure are $\mu_u^i = E(U_i|Z_i, W_i) = (I - \Gamma_1)^{-1}(\alpha + \Gamma_2 W_i)$ and $\Psi_u = \mathrm{var}(U_i|Z_i, W_i) = (I - \Gamma_1)^{-1}\Psi(I - \Gamma_1)^{-T}$; the mean for the surrogates is $\mu_x^i = E(X_i|Z_i, W_i) = \nu + \Lambda\mu_u^i + KW_i$; and the marginal variance of the surrogates is $\Omega_x = \mathrm{var}(X_i|Z_i, W_i) = \Lambda\Psi_u\Lambda^T + \Omega_\delta$. The conditional moments of the latent predictors, given the error prone measurements, are

$$(5) \qquad \widetilde{U}_i = E(U_i|X_i, Z_i, W_i) = \mu_u^i + \Psi_u\Lambda^T\Omega_x^{-1}(X_i - \mu_x^i),$$

$$(6) \qquad \widetilde{\Psi}_u = \mathrm{var}(U_i|X_i, Z_i, W_i) = \Psi_u - \Psi_u\Lambda^T\Omega_x^{-1}\Lambda\Psi_u.$$

Finally, for a subject with $n_i$ observed outcomes, the conditional moments of the outcome, given the error prone predictors, are functions of (5) and (6):

$$(7) \qquad \mu_{y|x}^{ij} = E(Y_{ij}|X_i, W_i, Z_i) = \beta_0 + \beta^T\widetilde{U}_i + \kappa^T Z_{ij},$$

$$(8) \qquad \Omega_{y|x} = \mathrm{var}(Y_i|X_i, W_i, Z_i) = \Omega_\varepsilon + \beta^T\widetilde{\Psi}_u\beta 1_{n_i}1_{n_i}^T,$$

where $1_{n_i}$ is a vector of ones of length $n_i$. For ease of exposition, in what follows we drop the subject index $i$, unless necessary.

In cases where the surrogate vector, $X$, is not completely observed, (5)–(8) depend on the missing data pattern for the given individual. Such dependence could be denoted by a subscript $(m)$ representing the $m$th missing data pattern among the $X$, for example, $\widetilde{U}_{(m)}, \widetilde{\Psi}_{u(m)}, \mu_{y|x(m)}$ and $\Omega_{y|x(m)}$. In particular, note that the variance (8) of the outcome depends on the missing data pattern for the surrogates, $X$. This dependence will play a role in developing weights for the proposed estimating equations approach.

4.1. *Full maximum likelihood estimation.* To estimate the model parameters via maximum likelihood, consider the likelihood contribution of one subject, conditional on covariates, $Z, W$:

$$L(\theta) = f(Y, X|Z, W; \theta) = f(Y|X, Z, W; \theta_1, \theta_2, \theta_3)f(X|Z, W; \theta_3).$$

Assuming normality of $\varepsilon$, $\delta$ and $\xi$, then $f(Y|X,Z,W;\theta_1,\theta_2,\theta_3) \sim$ Normal$(\mu_{y|x}, \Omega_{y|x})$, and $f(X|Z,W;\theta_3) \sim$ Normal$(\mu_x, \Omega_x)$. Thus, letting $\ell(\theta) = \log L(\theta)$, letting subscript $k$ denote the $k$th element of $\beta$, and letting Tr denote the trace of a matrix, a subject's contribution to the likelihood score equations for $\theta_1$ is given by

$$\frac{\partial \ell}{\partial \beta_0} = 1^T \Omega_{y|x}^{-1}(Y - \mu_{y|x}),$$

$$\frac{\partial \ell}{\partial \beta_k} = \widetilde{U}_k \Omega_{y|x}^{-1}(Y - \mu_{y|x})$$
$$+ \frac{1}{2} \text{Tr}\left\{\Omega_{y|x}^{-1} \frac{\partial \Omega_{y|x}}{\partial \beta_k} \Omega_{y|x}^{-1}[(Y - \mu_{y|x})(Y - \mu_{y|x})^T - \Omega_{y|x}]\right\},$$

$$\frac{\partial \ell}{\partial \kappa} = Z \Omega_{y|x}^{-1}(Y - \mu_{y|x}).$$

Similarly, letting the subscripts $2k$ and $3k$ denote the $k$th element of $\theta_2$ and $\theta_3$ respectively, the contributions for the outcome variance and the exposure model parameters are

$$\frac{\partial \ell}{\partial \theta_{2k}} = \frac{1}{2} \text{Tr}\left\{\Omega_{y|x}^{-1} \frac{\partial \Omega_{y|x}}{\partial \theta_{2k}} \Omega_{y|x}^{-1}[(Y - \mu_{y|x})(Y - \mu_{y|x})^T - \Omega_{y|x}]\right\},$$

$$\frac{\partial \ell}{\partial \theta_{3k}} = \frac{\partial}{\partial \theta_{3k}} \log(f(Y|X,Z,W;\theta_1,\theta_2,\theta_3)) + \frac{\partial}{\partial \theta_{3k}} \log(f(X|Z,W;\theta_3)).$$

Under correct model specification, setting the score equal to zero can be used to obtain asymptotically unbiased parameter estimates, $\widehat{\theta}_{\text{MLE}}$. The variance of $\widehat{\theta}_{\text{MLE}}$ can be calculated from the inverse of the information matrix, $\text{var}(\widehat{\theta}_{\text{MLE}}) = [-\sum_{i=1}^N E(\partial^2 \ell_i / \partial\theta \, \partial\theta^T)]^{-1}$. If normality assumptions are not satisfied, then robust standard error estimates can be computed [Arminger and Schoenberg (1989)].

In contrast to linear models for longitudinal data where covariates are measured without error [Laird and Ware (1982)], misspecification of the error covariance, $\Omega_\varepsilon$, of the longitudinal outcome may induce bias in the estimate for $\beta$. This is seen by examining the expected value of its score equation. The first term in $\partial \ell / \partial \beta_k$ is zero in expectation if the conditional mean $\mu_{y|x} = \beta_0 1_n + \beta^T \widetilde{U} 1_n + \kappa^T Z$ is correctly specified, which occurs when: (a) $X$ are surrogates [i.e., $f(Y|U,X,Z,W) = f(Y|U,Z,W)$, Carroll et al. (2006)]; (b) the variance for the error prone surrogates is correctly modeled [i.e., $\text{var}(X|W,Z) = \Lambda \Psi_u \Lambda^T + \Omega_\delta$]; and when (c) the variance for the latent predictors is correctly modeled, $\text{var}(U|Z,W) = \Psi_u$. The second term in $\partial \ell / \partial \beta_k$ is zero in expectation, assuming that the conditional variance of the outcome given the surrogates, $\Omega_{y|x} = \Omega_\varepsilon + \beta^T \widetilde{\Psi}_u \beta$, has been correctly specified. The conditional variance is correct only when the conditions above are met, in addition to (d) correctly modeling $\Omega_\varepsilon$, the variance of the longitudinal responses given the true exposure. In Section 5 we assess the magnitude of bias in $\widehat{\beta}$ induced due to misspecification of $\Omega_\varepsilon$.

4.2. *Estimating equations approach.* We formulate generalized estimating equations [Liang and Zeger (1986)] that relax the dependence on correct specification of the conditional error variance structure, $\Omega_\varepsilon$. The estimating equations are given by $\mathcal{S} \equiv \sum S_i = 0$, where $\mathcal{S} = (\mathcal{S}_{\theta_1}^T, \mathcal{S}_{\theta_2}^T, \mathcal{S}_{\theta_3}^T)^T$ has three components corresponding to $\theta_1, \theta_2, \theta_3$. Each subject's contribution, $S_i$, can be similarly partitioned. Letting the subscript $k$ represent the $k$th element of a vector, and dropping the subscript $i$, an individual's contribution to $\mathcal{S}$ is given by

$$S_{\beta_0} = R_{y|x}^{-1}(Y - \mu_{y|x}),$$
$$S_{\beta_k} = \widetilde{U}_k 1_n^T R_{y|x}^{-1}(Y - \mu_{y|x}),$$
$$S_\kappa = Z^T R_{y|x}^{-1}(Y - \mu_{y|x}),$$
$$S_{\theta_{2k}} = \frac{1}{2} \text{Tr}\left\{\left(R_{y|x}^{-1} \frac{\partial \Omega_{y|x}}{\partial \theta_{2k}} R_{y|x}^{-1}\right)((Y - \mu_{y|x})(Y - \mu_{y|x})^T - \Omega_{y|x})\right\},$$
$$S_{\theta_{3k}} = \frac{\partial}{\partial \theta_{3k}} \log(f(X|Z,W;\theta_3)),$$

where $R_{y|x}$ is a working covariance matrix discussed later. The estimating equations relax the dependence of correct specification of $\Omega_\varepsilon$ by eliminating the second term in the score equations for $\beta$. Further, note that $\mathcal{S}_{\theta_3}$ does not depend on $\theta_1$ or $\theta_2$, which makes the estimation of exposure model parameters, $\theta_3$, robust to outcome model misspecification. Also, it allows the equations for $\theta_3$ to be solved separately from those of $\theta_1$ and $\theta_2$. The estimates obtained from these estimating equations will be called $\widehat{\theta}_{\text{EE}}$.

The variance for the estimates obtained via the estimating equations can be calculated from the sandwich formula, $N\text{var}(\widehat{\theta}_{\text{EE}}) = B^{-1}AB^{-T}$. The components of the formula, $A = E(\mathcal{S}\mathcal{S})$ and $B = E(\partial \mathcal{S}/\partial \theta^T)$, can be consistently estimated by their empirical analogs

$$\widehat{B} = \frac{1}{N} \sum_{i=1}^N \left(\frac{\partial S_i}{\partial \theta^T}\bigg|_{\theta=\widehat{\theta}}\right) \quad \text{and} \quad \widehat{A} = \frac{1}{N} \sum_{i=1}^N (S_i S_i^T)|_{\theta=\widehat{\theta}}.$$

As detailed in the supplementary materials [Sánchez, Budtz-Jørgensen and Ryan (2009a)], $A$ and $B$ have a block structure, which enables us to write the following variance formula for $\widehat{\theta}_1$:

(9)   $\text{var}(\widehat{\theta}_{1,\text{EE}}) = B_{11}^{-1} A_{11} B_{11}^{-T} + B_{11}^{-1} B_{13} \text{var}(\widehat{\theta}_{3,\text{EE}}) B_{13}^T B_{11}^{-T}.$

The first term in (9), $B_{11}^{-1} A_{11} B_{11}^{-T}$, is the asymptotic variance of $\widehat{\theta}_1$ obtained from naively regressing $Y$ on the estimated exposure, $\widetilde{U}$ (evaluated at $\widehat{\theta}_{3,\text{EE}}$). The second term in (9) adjusts the variance of the naive estimator by accounting for the estimation of the exposure model parameters, and $\text{var}(\widehat{\theta}_{3,\text{EE}}) = B_{33}^{-1} A_{33} B_{33}^{-T}$.

We study three alternate forms for the working covariance matrix, $R_{y|x}$. First, we set $R_{y|x} = \Omega_{y|x} = \Omega_\varepsilon + \beta^T \widetilde{\Psi}_u \beta 11^T$, with $\widetilde{\Psi}_u$ evaluated at $\widehat{\theta}_3$. Except for the fact that the estimate of $\theta_3$ is slightly different, this working covariance is identical to the weight matrix of the score equations. Further, because $R_{y|x}$ depends on $\widetilde{\Psi}_u$, subjects with missing data among the surrogates $X$ are differentially weighted during the estimation. We call the estimates from this approach $\widehat{\beta}_{\text{EE1}}$.

The other two working covariance matrices are $R_{y|x} = \Omega_\varepsilon$, and $R_{y|x} = \Omega_\varepsilon + \beta_*^T \widetilde{\Psi}_u \beta_* 11^T$, where $\beta_*$ is a fixed, "best guess" value of $\beta$. The former is a special case of the latter with $\beta_* = 0$. Neither of these matrices depend on the unknown exposure effect $\beta$. We will call estimates obtained using the working correlation $R_{y|x} = \Omega_\varepsilon + \beta_*^T \widetilde{\Psi}_u \beta_* 11^T$ with $\beta_* \neq 0$ as $\widehat{\beta}_{\text{EE2}}$. When $\beta_* = 0$, solving $(\mathcal{S}_{\theta_1}^T, \mathcal{S}_{\theta_2}^T)^T = 0$ corresponds to fitting the outcome model with the true exposure $U$ replaced by the estimate $\widetilde{U}$ (evaluated at $\widehat{\theta}_{3,\text{EE}}$) and a working covariance that is independent of the exposure model. As detailed below, this last procedure is similar to regression on factor scores and regression calibration [Carroll et al. (2006)]. This estimate will be denoted $\widehat{\beta}_{\text{RC}}$. The three estimating equation estimators under study are as follows: $\widehat{\beta}_{\text{EE1}}$, $\widehat{\beta}_{\text{EE2}}$ and $\widehat{\beta}_{\text{RC}}$.

4.3. *Connections to regression on factor scores and regression calibration.* The proposed estimation approach can be compared and contrasted to regression on factor scores, which has been widely used in the social sciences to estimate models with one or more latent variables. It consists of two steps. First, factor analysis is used to derive estimates of the latent variables called *factor scores*. At least two ways of estimating factor scores exist [e.g., Skrondal and Laake (2001)]. Tucker (1971) recommends empirical Bayes (EB) estimation of factor scores when they are to be used as predictor variables in a regression model. Thus, estimating the exposure model in (2)–(3), and using $\widetilde{U}$ evaluated at $\widehat{\theta}_{3,\text{EE}}$ as the estimated factors completes the first step of regression on factor scores, with latent variables estimated by EB estimation. The second step is to use the estimated factors as predictors (in our case) or outcomes in a regression model. This would correspond to first estimating $\theta_3$ from $\mathcal{S}_{\theta_3}^T = 0$, and then plugging in the resulting estimate when solving for $\theta_1$ and $\theta_2$. Thus, estimating the exposure effect using regression on factor scores corresponds to obtaining $\widehat{\beta}_{\text{RC}}$.

Factor score regression, however, has stopped short in that correct standard errors for regression parameters are not computed. That is, the standard errors for regression parameters in factor score regression do not typically account for the estimation uncertainty of $\widehat{\theta}_{3,\text{EE}}$. Recently, resampling approaches were suggested as a means to correct such standard errors [Skrondal and Laake (2001)]. The proposed estimating equations formalize the ad-hoc regression on factor scores procedure by expressing the procedure in an estimating equations framework, and thereby allowing direct estimation of correct standard errors without the need for resampling techniques. Further, by introducing $R_{y|x}$ as weights, weighed factor score regression is made possible, where observations are weighed according to the amount of exposure information available.

Some researchers argue that regression parameters estimated via factor score regression are biased [e.g., Bollen (1989); Wall and Li (2003)] because the estimated factors $\widetilde{U}$ are still error prone measures of the true latent variables. As documented by Skrondal and Laake (2001), the biases can be avoided by carefully choosing the way factor scores are estimated (e.g., EB estimation when latent variables are predictors). In the proposed framework, it is easy to see that $\mathcal{S}_\beta$ has expectation zero such that $\widehat{\beta}_{\text{EE}}$ is, at least asymptotically, unbiased, assuming the exposure model is correctly specified. Parallel arguments to those described in measurement error methodology can also be made for the unbiasedness of $\widehat{\beta}_{\text{EE}}$. Namely, the measurement error of $\widetilde{U}$ is of Berkson-type [i.e., the error is independent of $\widetilde{U}$: $E(U_i - \widetilde{U}_i|\widetilde{U}_i, Z_i, W_i) = E(U_i|X_i, Z_i, W_i) - \widetilde{U}_i = 0$], which, in the linear model, inflates the standard error estimates for $\beta$ but does not introduce bias [Fuller (1987); Carroll et al. (2006)].

The proposed estimating equations are philosophically similar to regression calibration, although some differences exist. Regression calibration consists of regressing the outcome on the estimated exposure given only the surrogates and covariates [i.e., using $\widetilde{U}_i = E(U_i|X_i, Z_i, W_i)$ as a predictor], and subsequently obtaining correct standard errors for the regression parameters. In the proposed approach, an estimated value of the latent exposure which depends only on the surrogates and covariates can be obtained by first using $\mathcal{S}_{\theta_3} = 0$ to solve for $\widehat{\theta}_3$, and then calculating an estimated exposure using (5) evaluated at $\widehat{\theta}_3$. Next, similar to regression calibration, the estimated exposure can be plugged into $\mathcal{S}_{\theta_1} = 0$ and $\mathcal{S}_{\theta_2} = 0$ to solve for $\widehat{\theta}_1$ and $\widehat{\theta}_2$. In regression calibration, however, assumptions on the distribution of the surrogates (distribution of measurement error) are not made, other than having mean zero conditional on observed covariates. Exposure model parameters, $\theta_3$, are often estimated via method of moments. In contrast, $\mathcal{S}_{\theta_3}$ is derived from normality assumptions, which were necessary to easily account for surrogates missing at random [Little and Rubin (2002)]. That is, since exposure measurements $X$ may often be missing not completely at random and many patterns of missing data are possible, we opt for maximum likelihood as a method to estimate $\theta_3$. Last, regression calibration sets $R_{y|x} = \Omega_\varepsilon$; that is, it eliminates the dependence of $R_{y|x}$ on the missing pattern for $X$. Thus, regression calibration does not provide differential weighting for observations with more or less exposure information.

4.4. *A note on the practical implementation of estimation approaches.* Many software packages are available to estimate latent variable models using maximum likelihood [e.g., Jöreskog and Sörbom (1989); Muthén and Muthén (1998–2004); Rabe-Hesketh et al. (2004); Fox (2006)]. A few can accommodate estimation when data is missing at random, and/or when the variable's distribution is not normal [e.g., Muthén and Muthén (1998–2004); Rabe-Hesketh et al. (2004)]. A more complete software review is available [Rabe-Hesketh and Skrondal (2008)].

In the example and in the simulations presented in the next section we used the package Mplus to obtain maximum likelihood estimates. To obtain the estimating equation estimates we used a combination of Mplus [Muthén and Muthén (1998–2004)] and R. Because the estimating equations for $\theta_3$ can be solved separately from those of $\theta_1, \theta_2$, we first used Mplus to solve $\mathcal{S}_{\theta_3} = 0$. Estimates of $\widetilde{U}$ were obtained in R by importing the data and Mplus parameter estimates. Subsequently, we used R to solve $\mathcal{S}_{\theta_1}(\widehat{\theta}_3) = 0, \mathcal{S}_{\theta_2}(\widehat{\theta}_3) = 0$. Finally, corrected standard errors derived from (9) were obtained. All functions in R used to solve for parameter estimates and compute corrected standard errors are available as supplementary materials from the journal's website [Sánchez, Budtz-Jørgensen and Ryan (2009b)].

## 5. Comparing estimation approaches: bias, mean squared error and efficiency.

5.1. *Bias and mean squared error under a misspecified model.* Via a simulation study, we assessed the magnitude of the bias and mean squared error of the estimated exposure effect $\widehat{\beta}$ when the conditional covariance structure, $\Omega_\varepsilon$, is misspecified. Data were generated from a one latent variable model with longitudinal outcomes, where $\Omega_\varepsilon$ was designed to have a heterogeneous autocorrelation structure with parameter $\rho$. We considered a range of values of $\beta$, and various strengths of the correlation in the outcomes, $\rho = 0, 0.25, 0.5, 0.75$ ($\rho = 0$ allows us to assess the effect of incorrectly assuming equal variance across time). Other parameters are detailed in the supplementary materials [Sánchez et al. (2009a)]. For each combination of $\beta$ and $\rho$, two hundred data sets of $N = 500$ were generated. An otherwise correctly specified model was then fitted to the simulated data, except that independence or compound symmetry, $\Omega_\varepsilon = \sigma^2 I + \sigma_w^2 11^T$, variance structures were assumed. Data for the surrogates of exposure was complete.

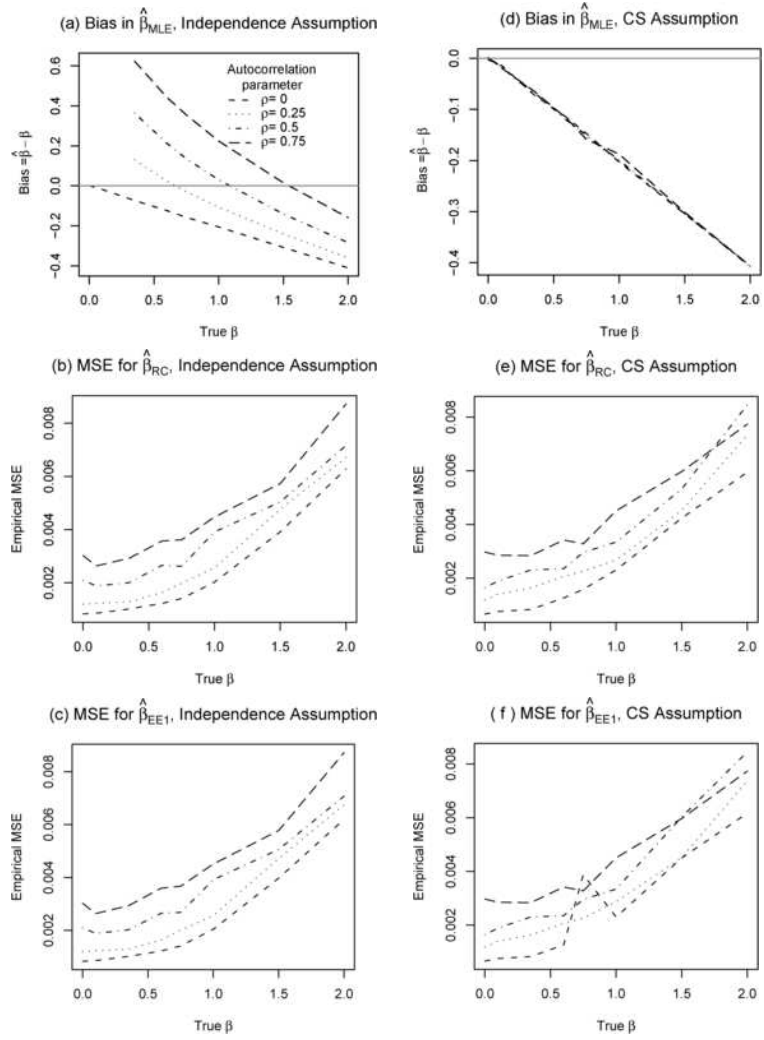

Fig. 2. *Bias and Mean Square Error for various estimation approaches, under incorrect conditional covariance assumption for the longitudinal outcome. In the true model, $\Omega_\varepsilon$ has a heterogeneous autoregressive structure of order 1 with various strengths of the autocorrelation parameter $\rho$ as shown in panel* (a). *Panels* (a)–(c) *and* (d)–(e) *show the resultant bias or MSE for* $\widehat{\beta}_{\mathrm{MLE}}$, $\widehat{\beta}_{\mathrm{RC}}$ *and* $\widehat{\beta}_{\mathrm{EE1}}$ *under an incorrect independence and compound symmetry assumption in the fitted model, respectively. For* $\widehat{\beta}_{\mathrm{MLE}}$ *the bias dominates the bias; for* $\widetilde{\beta}_{\mathrm{RC}}$ *and* $\widetilde{\beta}_{\mathrm{EE1}}$ *the bias is a negligible fraction of the MSE.*

Figure 2(a) displays the average bias for the exposure effect estimated via maximum likelihood when an independence variance structure is incorrectly assumed. The bias in $\widehat{\beta}_{\text{MLE}}$ can be either positive or negative. The bias is positive when the true exposure effect is small, but is negative for larger effects. Correctly assuming conditional independence but incorrectly assuming homogenous variance results, in relative terms, in underestimation of the exposure effect of about 20% [Figure 2(a), short-dash curve]. In contrast, the bias for the regression calibration estimator and the proposed estimating equations ($\widehat{\beta}_{\text{EE1}}$) is nearly zero in all scenarios (bias not shown). Figures 2(b) and (c) show the empirical mean squared error for these estimators, of which the bias is a negligible part. The bias for $\widehat{\beta}_{\text{EE2}}$ is similar to that of $\widehat{\beta}_{\text{RC}}$ and $\widehat{\beta}_{\text{EE1}}$, independent of the value of $\beta_*$ (figure not shown). The mean squared error for the regression calibration estimate and that of $\widehat{\beta}_{\text{EE1}}$ are approximately the same, and are much smaller than that of $\widehat{\beta}_{\text{MLE}}$.

Figures 2(d)–(f) show the bias or mean squared error resulting from incorrectly assuming a compound symmetry structure. Here, the MLE estimates are biased toward the null, with the magnitude of the bias increasing with increasing magnitude of the true effect. In relative terms the bias is about 20% irrespective of the true effect or the correlation parameter $\rho$. Again the bias in the estimates obtained via estimating equations is nearly zero, and the mean squared error for the regression calibration estimate was the same as that of $\widehat{\beta}_{\text{EE1}}$.

5.2. *Theoretical relative efficiency under correct model specification: $\widehat{\theta}_{\text{EE1}}$ vs $\widehat{\theta}_{\text{MLE}}$.* We consider the efficiency of $\widehat{\theta}_{\text{EE1}}$ (i.e., using weights that depend on the unknown $\beta$) relative to $\widehat{\theta}_{\text{MLE}}$ by studying the information matrix of both estimators, and the form of the conditional variance $\Omega_\varepsilon = \text{var}(Y|U, Z, W)$. As detailed in the supplementary materials [Sánchez et al. (2009a)], the similarity in the form of the information matrices provides an insight as to when the estimating equations approach loses information compared to maximum likelihood. When $\Omega_\varepsilon$ is parameterized with linear functions of $\theta_2$ (e.g., compound symmetry, unstructured, banded), it can be shown that the estimating equations approach loses information only in estimating $\theta_3$. However, because $\widehat{\theta}_1$ is not independent of $\widehat{\theta}_3$, the loss of information in $\widehat{\theta}_3$ affects the asymptotic relative efficiency of $\widehat{\theta}_1$. When $\Omega_\varepsilon$ is parameterized with nonlinear functions of $\theta_2$ (e.g., autoregressive structure), information is also lost for $\widehat{\theta}_1$, and higher losses in efficiency are expected.

To quantify the loss of efficiency in the parameter of interest, $\beta$, we evaluated exact expressions for the relative efficiency of $\widehat{\beta}_{\text{EE1}}$ compared to $\widehat{\beta}_{\text{MLE}}$, again using a model with one latent exposure and longitudinal outcomes as in Section 5.1. We considered a compound symmetry structure for $\varepsilon$'s variance, $\Omega_\varepsilon = \sigma^2 I + \sigma_w^2 1 1^T$, in addition to an autoregressive structure of order one. In the calculations we allowed $\beta$ to vary over a range of values, and considered three scenarios of the magnitude of the surrogates' measurement error variances. The specific parameter values for the simulations are given in the supplementary materials [Sánchez et al. (2009a)].

Figure 3 shows the results of the efficiency calculations, and illustrates the dependency of the relative efficiency on several parameters. First, the loss of efficiency in $\widehat{\beta}_{\text{EE1}}$ is fairly small for small values of the exposure effect, but increases with larger values of $\beta$. This increase can be explained by the fact that when the exposure effect $\beta$ increases, the outcome model will hold more information about the exposure model parameters $\theta_3$. This information is utilized in MLE, but not in the estimating equations. The loss of information in $\widehat{\theta}_3$ affects the efficiency of $\widehat{\beta}$ because these parameters are not independent. Efficiency also depends on the structure of $\Omega_\varepsilon = \text{var}(Y|U, Z, W)$. Under compound symmetry, Figure 3(a)–(c), the relative efficiency does not exceed 1.06. In contrast, the loss of efficiency under an autoregressive structure for the $\Omega_\varepsilon$ can range up to 25% for large effects, Figure 3(d)–(f). The larger loss of efficiency under the autoregressive structure is to be expected given that the covariance matrix is not linear in $\rho$, as previously noted. Under a given correlation structure, the loss of efficiency decreases with stronger correlations in the outcome. Finally, the loss of efficiency increases when the surrogates' measurement error variances increase. In applications this may be relevant when selecting between maximum likelihood and the estimating equations approach in the presence of poor exposure surrogates.

5.3. *Variance ratios under correct model specification: $\widehat{\beta}_{\text{EE2}}$ and $\widehat{\beta}_{\text{RC}}$ vs $\widehat{\beta}_{\text{EE1}}$.* We compared the variance of $\widehat{\beta}_{\text{EE2}}$ and $\widehat{\beta}_{\text{RC}}$ to that of $\widehat{\beta}_{\text{EE1}}$; that is, we evaluated the effect of replacing $R_{y|x} = \Omega_\varepsilon + \beta^T \widetilde{\Psi}_u \beta 1_n 1_n^T$, with $R_{y|x} = \Omega_\varepsilon + \beta_*^T \widetilde{\Psi}_u \beta_* 1_n 1_n^T$, where $\beta_*$ is a fixed constant (including zero). To calculate variance ratios, we simulated data from a model with one latent exposure, similar to that described in Section 5.1. We considered, in addition, a larger number of exposure surrogates $p = 12$, and allow surrogate data to be missing completely at random under various scenarios. In the first scenario, missing data patterns were equally likely. In a second scenario the probability of a missing data pattern was proportional to the theoretical variance of the latent variable given the observed pattern of surrogates, $\text{var}(U|X_{(m)}, Z, W)$. In the third, the probability of a missing data pattern was inversely proportional to $\text{var}(U|X_{(m)}, Z, W)$. Hence, in the first scenario the distribution of the values of $\text{var}(U|X_{(m)}, Z, W)$ have a uniform distribution, whereas in the second and third, these variances were skewed to the left and right, respectively. For each of the three values of $\beta = 0.5, 1, 2$, one thousand data sets of $N = 1000$ observations were simulated. For each data set, we obtained $\widehat{\beta}_{\text{EE1}}$, and for a range of values of fixed $\beta_*$, we obtained $\widehat{\beta}_{\text{EE2}}$ and $\widehat{\beta}_{\text{RC}}$ ($\beta_* = 0$). The empirical variances of the estimated effects were then calculated as well as their ratios.

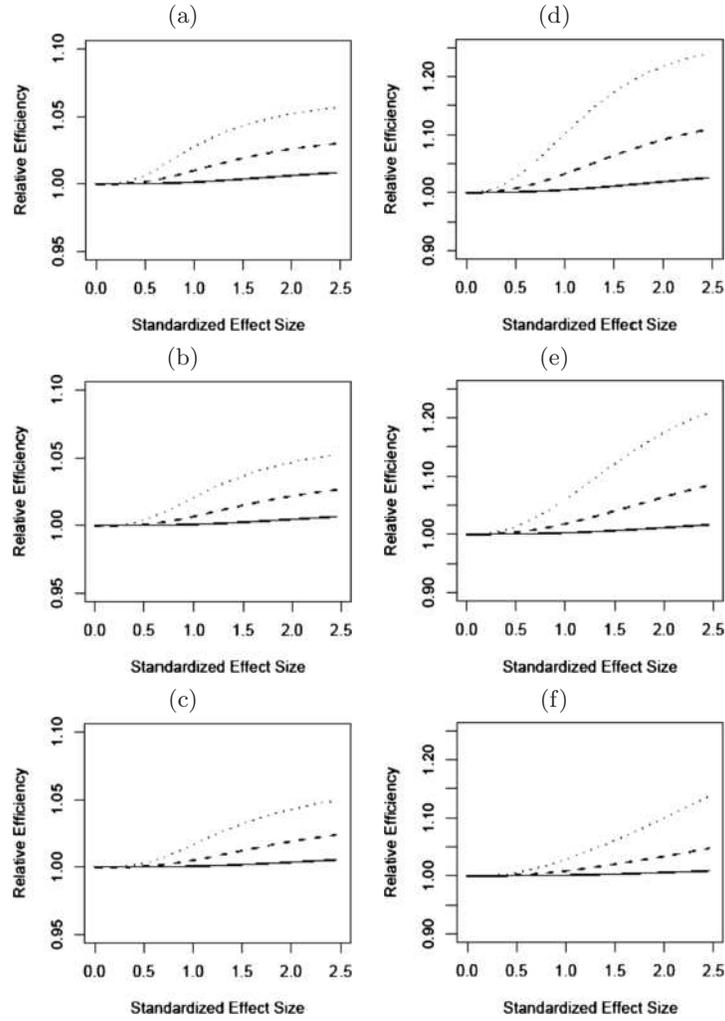

FIG. 3. *Relative efficiency of maximum likelihood estimate $(\widehat{\beta}_{\mathrm{MLE}})$ compared to estimating equations approach $(\widehat{\beta}_{\mathrm{EE1}})$ in the case of compound symmetry error structure with* (a) $\sigma_w^2/(\sigma_w^2 + \sigma^2) = 0.25$; (b) $\sigma_w^2/(\sigma_w^2 + \sigma^2) = 0.50$; *and* (c) $\sigma_w^2/(\sigma_w^2 + \sigma^2) = 0.75$; *and autoregressive error structure with* (d) $\rho = 0.25$; (e) $\rho = 0.50$; *and* (f) $\rho = 0.75$. *Plots show a range of values of the exposure effect $\beta$ in standardized units [standardized($\beta$) $= \beta\sqrt{\mathrm{var}(U)}/\sqrt{\sigma^2 + \sigma_w^2}$]. Curves represent varying degrees of measurement error in the surrogates, expressed as percentage of variability in measurement: 10% (solid), 22% (dashed), 36% (dotted).*

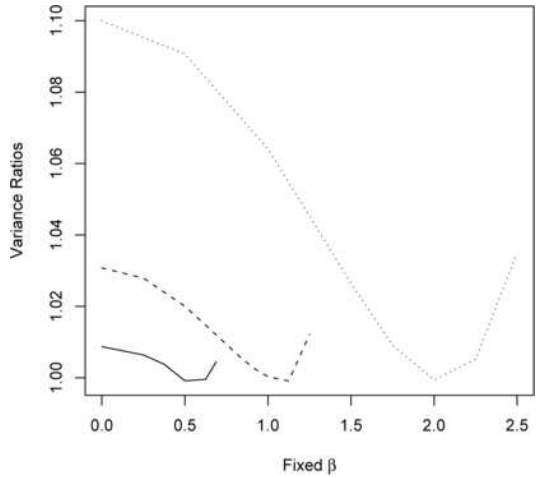

FIG. 4. *Figure compares the variance for $\widehat{\beta}_{\mathrm{EE2}}$ to the variance of $\widehat{\beta}_{\mathrm{EE1}}$ for three values of the true effect: $\beta = 0.5$ (solid), $\beta = 1$ (dashed), $\beta = 2$ (dotted). The ratio is shown as a function of the fixed value of $\beta$ used to obtain $\widehat{\beta}_{\mathrm{EE2}}$. Note that Fixed $\beta = 0$ corresponds to $\widehat{\beta}_{\mathrm{RC}}$.*

Figure 4 shows the simulation results for the case when the values of $\mathrm{var}(U|X_{(m)}, Z, W)$ have a distribution that is skewed to the left and the number of surrogates is 12. From the figure, we see that utilizing a fixed value of $\beta$, $\beta_*$, in $R_{y|x} = \Omega_\varepsilon + \beta_*^T \widetilde{\Psi}_u \beta_* 1_n 1_n^T$ is practically as good as estimating $\beta$ in the weights when $\beta_*$ is close to the truth. Further, we see that efficiency gains of up to 10% are obtained by using a fixed value of $\beta_*$ close to the truth in comparison to a regression calibration ($\beta_* = 0$). For the case when the missingness probability was equally likely or inversely proportional to $\mathrm{var}(U|X_{(m)}, Z, W)$, the maximum variance ratio observed from a figure similar to Figure 4 was more modest at 2%. Similar patterns were observed when the number of surrogates was $p = 3$.

5.4. *Remarks on fetal lead exposure example.* Substantial differences in the estimated variances of the exposure effect were not observed for the different estimating equations estimates (Table 2). A reason is that although there is a large number of missing data patterns for the exposure measurements, most subjects have enough surrogates to make the variance of the estimated exposure relatively small and similar across subjects. Further, although the estimated exposure effect is a 1.0 point decline in MDI scores for a doubling in plasma lead concentration at trimester 1, in standardized units the effect is roughly 0.19. Thus, given the results of our simulation study, considerable differences in efficiency among the estimating equations approaches are not expected.

**6. Conclusions.** We used generalized estimating equations to estimate parameters in a model for longitudinal responses where the predictors are latent. We were motivated by large potential biases in the maximum likelihood estimate when the conditional variance of the outcome given the latent exposure is misspecified. The estimating equations approach relaxes the assumption of correct conditional variance specification for the longitudinal outcomes. When the model is correctly specified, the loss of efficiency of the estimating equations is small for small effect sizes. When weighed estimating equations are used and the exposure effect estimated as part of the weights, the loss of efficiency is negligible. When the exposure measurements (surrogates) are missing, the relative efficiency depends on the distribution of the conditional variance of the latent exposure given observed surrogates (these variances depend on the missing data pattern). Our estimation approach encompasses regression calibration and regression on factor scores as special cases.

The estimating equations approach enables us to show that regression on factor scores yields asymptotically unbiased effect estimates when the exposure model is correctly specified, and allows us to provide a conceptually simple way of computing correct standard errors for the outcome model parameters without the need for resampling approaches. In the setting of factor score regression, Skrondal and Laake (2001) and Tucker (1971) discussed a sufficient requirement to obtain unbiased regression parameters, namely, estimating the factor scores using an Empirical Bayes approach we also discussed when latent variables are predictors. However, they did not provide a simple way of conducting inference on the regression parameters. Further, they did not consider studies with longitudinal outcomes, which have the additional subtlety of misspecified residual error covariance structures.

Sample size is an important consideration when justifying the use of complex latent variable models. Rules of thumb in the literature recommend anywhere between five to twenty times cases as there are variables in the model to ensure stable parameter estimates [Bentler and Chou (1987); Stevens (1996)]. To fit the exposure model in the example, at least 260 observations would be needed. Sample sizes may also impact the bias and variance comparison results in this paper. While a full evaluation of the impact of sample size on our results is beyond the scope of this paper, we conducted some exploratory simulations with sample sizes as low as 250. At this reduced sample size, the bias of the MLE was as strong as that portrayed in Figure 2, while the bias of the estimating equations remained very small. The efficiency loss in $\widehat{\beta}_{\text{EE1}}$ compared to $\widehat{\beta}_{\text{MLE}}$ also remained low, for example, the highest value in a figure similar to Figure 3(c) was 1.03. Finally, the variance ratio comparing $\widehat{\beta}_{\text{RC}}$ to $\widehat{\beta}_{\text{EE1}}$ was 1.12 when the true effect was 2 (i.e., about the same value as the highest point in Figure 4). Further examination of sample size issues may be warranted, particulary for cases of small sample size and missing data among surrogates.

Practical advantages and disadvantages of the proposed method are as follows. The estimating approach enables the practice of reusing stored estimates of latent exposure values in analyses of various health outcomes. This is advantageous in large epidemiological studies where testing the effects of a latent exposure on several health outcomes might be of interest. In such scenarios, correctly estimating effect's variances requires computing derivatives of the estimating equations for each outcome's mean parameters (i.e., obtaining $B_{13}$). Automated procedures for this computation can be easily implemented. However, separately estimating the exposure model might give rise to identifiability problems in cases where there are only two surrogates for a latent variable. Such identifiability problems might not arise with full maximum likelihood estimation. Furthermore, this approach is confined to continuous outcomes. We do not expect the unbiasedness of the approach to hold for other outcomes, for which modifications may be needed [Carroll et al. (2006)]. Last, the approach still requires correct specification of the conditional error variance structure in the exposure part of the model. The more difficult problem of relaxing correct specification of the exposure model variance structure remains to be studied.

## SUPPLEMENTARY MATERIAL

**Supplement A: Supplement to "An estimating equations approach to fitting latent exposure models with longitudinal responses"** (DOI: 10.1214/08-AOAS226SUPPA; .pdf). We provide details on the simulation parameters referenced in Sections 5.1, 5.2 and 5.3. We also provide details on variance and relative efficiency calculations mentioned in Sections 4.2 and 5.2, respectively.

**Supplement B: Computer code supplement to "An estimating equations approach to fitting latent exposure models with longitudinal responses"** (DOI: 10.1214/08-AOAS226SUPPB; .zip). This zipped folder contains several files with example code for the procedures described in this article. For specific details on the files, read the "readme.txt" file found within this zipped folder.

**Acknowledgments.** The authors would like to thank Professor James H. Ware and Professor Howard Hu for valuable discussions and the use of the data for this article. It has not been formally reviewed by the ACC or HHMI, and the views expressed in this document are solely those of the authors.

B. N. Sánchez  
Department of Biostatistics  
University of Michigan  
Ann Arbor, Michigan 48109  
USA  
E-mail: brisa@umich.edu

E. Budtz-Jørgensen  
Department of Biostatistics  
University of Copenhagen  
1014 Copenhagen  
Denmark  
E-mail: E.Budtz-Joergensen@biostat.ku.dk

L. M. Ryan  
Department of Biostatistics  
Harvard School of Public Health  
Boston, Massachusetts 02115  
USA  
E-mail: lryan@hsph.harvard.edu